**Different scenarios in the Dynamics of SARS-Cov-2 Infection: an adapted ODE model.**

Ramón E. R. González

*Departamento de Física, Universidade Federal Rural de Pernambuco, Recife, Pernambuco, CEP 52171-900, Brazil.*

**Abstract**

A mathematical model to calculate the transmissibility of SARS-Cov-2 in Wuhan City was developed and published recently by Tian-Mu Chen et al., Infectious Diseases of Poverty, 2020, https://doi.org/10.1186/s40249-020-00640-3. This paper improves this model in order to study the effect of different scenarios that include actions to contain the pandemic, such as isolation and quarantine of infected and at-risk people. Comparisons made between the different scenarios show that the progress of the infection is found to strongly depend on measures taken in each case. The particular case of Brazil was studied, showing the dynamics of the first days of the infection in comparison with the different scenarios contained in the model and the reality of the Brazilian health system was exposed, in front of each possible scenario. The relative evolution of the number of new infection and reported cases was employed to estimate a containment date of the pandemic. Finally, the basic reproduction number $R_0$ values were estimated for each scenario, ranging from 4.04 to 1.12.

## 1. Introduction

On February 26, the first case of SARS-Cov-2 was reported in Brazil [2]. From that day on, and taking into account the spread of the epidemic in China and Europe, in Brazil and in many other countries, a group of measures began to be taken, which, with their particularities in each country or region, aims to help minimize the effects of what was later classified by the World Health Organisation (WHO) as a global pandemic. The scientific community, playing its role in the midst of this global emergency, has focused its efforts on studying, among other things, the dynamics of the spread of the virus in several countries. Several studies disseminating models that study the dynamics of the epidemic have been carried out and published in these months from the first reported case [1, 3, 4, 5, 6]. One of the first was published by Tian-Mu Chen et al., Infectious Diseases of Poverty, 2020 [1]. In this work, the Bats-Hosts-Reservoir-People transmission network model was developed for simulating the potential transmission from the infection source (probably bats) to humans. We develop a new version of the model that only takes into account the transmission from person to person, as this is the dynamic for the rest of the countries except China, where the epidemic arose. The new version is a simple model of the SIR type, which includes some characteristic variables of the containment measures that have been taken worldwide. Parameters that define the proposed model are based on either the specific characteristics of the virus or the particularities of studied. Notice that, despite the dynamics of the epidemic shows a similar qualitative behaviour in any country, the specific characteristics of each region play a relevant role in the simulation process. The basic reproduction rate $R_0$ is defined from the parameters of the model. The model is applied to different scenarios, defined by different degree of adherence to the measures necessary to contain the spread of the epidemic. In order to demonstrate the importance of measures to contain the pandemic in the following topics, comparisons of the dynamics of infection and the spread of the virus are made according to the variables that represent these measures. The studies carried out are based on morphological characteristics of the infection curves such as the height of the peak of infection, its position and its width. Within each simulated scenario, the basic reproduction rate is calculated. Ro values are used to compare the different scenarios in order to define which is the most promising.

## 2. Adapted model based on Ordinary Differential Equations (ODE).

For countries where the virus was "imported", the original model is modified to exclude the host and the vector, taking into account the dynamics from several people initially infected with transmission capacity. To better illustrate the interactions between model variables, we can use the following scheme.

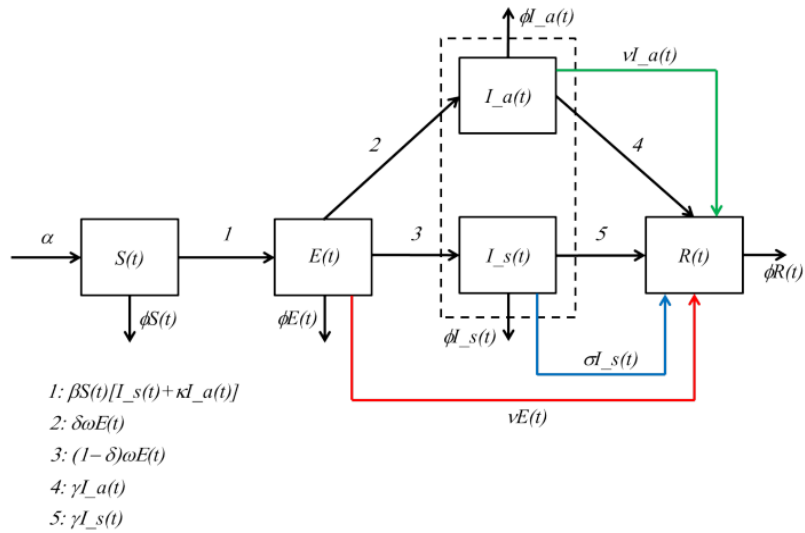

Figure 1. Representative scheme of the interactions between the different variables of the model.

The previous scheme shows the interactions between all variables from deterministic rules. These rules determine the stages through which individuals go through the entire process. The system of differential equations that describe these interactions, without including isolation and quarantine actions, is as follows (scenario 0):

$$\dot{S}(t) = \alpha - \phi S(t) - \beta S(t)[I\_s(t) + \kappa I\_a(t)] \qquad (1)$$

$$\dot{E}(t) = \beta S(t)[I\_s(t) + \kappa I\_a(t)] - (1-\delta)\omega E(t) - \delta\omega E(t) - \phi E(t) \qquad (2)$$

$$\dot{I}\_s(t) = (1-\delta)\omega E(t) - \gamma I\_s(t) - \phi I\_s(t) \qquad (3)$$

$$\dot{I}\_a(t) = \delta\omega E(t) - \gamma I\_a(t) - \phi I\_a(t) \qquad (4)$$

$$\dot{R}(t) = \gamma I_{s(t)} + \gamma I\_a(t) - \phi R(t) \qquad (5)$$

In the model represented here, there are three groups of individuals: susceptible and exposed (S and E), infected symptomatic and asymptomatic (I_s and I_a) and recovered and/or killed (R). Susceptible individuals are exposed due to their interaction with infected individuals of both types, with different degree of exposure. Infection of symptomatic patients occur from exposed individuals, after the virus incubation period, with a certain probability; asymptomatic patients, on the other hand, happen from individuals who have been exposed and who have latent viral particles in their body. A percentage of all individuals who have been infected will be recovered and the removal of individuals of each type happens naturally and is related, in addition to death, to the "exit" of the system.

The virus incubation period reported in [1] is 5.2 days. As in the original model, we assume that this is equally your latency period, so $\omega = 0.1923$. We also use the infection period value reported by the original study, so $\gamma = 0.1724$. The values of the proportion of asymptomatic infection, $\delta = 0.5$ and the multiple of the transmissibility of I_a to I_s, $\kappa = 0.5$ are the same as in the original model. The removal rate depends on each country. In the case of our model, based on Brazil, it was calculated following the same methodology adopted by the authors of [1] and from data extracted from reliable sources [7], resulting in a value of $\phi = \sim 10^{-5}$. The transmission rate in our model is considered a variable parameter since it is directly related to health actions to be undertaken by the population and the state. The previous system of equations (Eqs. 1 - 5) indicates the dynamics for the most pessimistic scenario, where containment measures against the advance of the epidemic such as quarantine or population isolation are not adopted (scenario 0). The inclusion in our model of actions to contain the epidemic can be seen in the scheme of figure 1. In this scheme, these measures are represented with the following colours: blue, quarantine of symptomatic infected, red, isolation of exposed individuals and green, the isolation of asymptomatic infected people. The system considering only the quarantine of the symptomatic infected represents scenario 1, the system including scenario 1 plus the isolation of exposed individuals represents scenario 2 and the inclusion, in addition to the other two scenarios, of the isolation of asymptomatic individuals, represents scenario 3, that would be the most optimistic scenario, but less realistic.

Taking into account the scenarios described above, equations (2 - 5) are for scenario 1, where a term concerning the quarantine of symptomatic individuals, $\sigma I\_s(t)$ is included:

$$\dot{E}(t) = \beta S(t)[I\_s(t) + \kappa I\_a(t)] - (1-\delta)\omega E(t) - \delta\omega E(t) - \phi E(t) \quad (6)$$

$$\dot{I}\_s(t) = (1-\delta)\omega E(t) - \gamma I\_s(t) - \phi\sigma I\_s(t) - \sigma I\_s(t) \quad (7)$$

$$\dot{I}\_a(t) = \delta\omega E(t) - \gamma I\_a(t) - \phi I\_a(t) \quad (8)$$

$$\dot{R}(t) = \gamma I\_s(t) + \gamma I\_a(t) - \phi R(t) + \sigma I\_s(t) \quad (9)$$

For scenario 2 (inclusion of a term concerning the isolation of exposed individuals, $\nu E(t)$):

$$\dot{E}(t) = \beta S(t)[I\_s(t) + \kappa I\_a(t)] - (1-\delta)\omega E(t) - \delta\omega E(t) - \phi E(t) - \nu E(t) \quad (10)$$

$$\dot{I}\_s(t) = (1-\delta)\omega E(t) - \gamma I\_s(t) - \phi I\_s(t) - \sigma I\_s(t) \quad (11)$$

$$\dot{I}\_a(t) = \delta\omega E(t) - \gamma I\_a(t) - \phi I\_a(t) \quad (12)$$

$$\dot{R}(t) = \gamma I\_s(t) + \gamma I\_a(t) - \phi R(t) + \sigma I\_s(t) + \nu E(t) \quad (13)$$

And for scenario 3 (inclusion of a term concerning the isolation of asymptomatic individuals, $\nu I\_a(t)$):

$$\dot{E}(t) = \beta S(t)[I\_s(t) + \kappa I\_a(t)] - (1-\delta)\omega E(t) - \delta\omega E(t) - \phi E(t) - \nu E(t) \quad (14)$$

$$\dot{I}\_s(t) = (1-\delta)\omega E(t) - \gamma I\_s(t) - \phi I\_s(t) - \sigma I\_s(t) \quad (15)$$

$$\dot{I}\_a(t) = \delta\omega E(t) - \gamma I\_a(t) - \phi I\_a(t) - \nu I\_a(t) \qquad (16)$$

$$\dot{R}(t) = \gamma I\_s(t) + \gamma I\_a(t) - \phi R(t) + \sigma I\_s(t) + \nu E(t) + \nu I\_a(t) \qquad (17)$$

The following table (Table 1) lists the different parameters and the initial conditions used in the simulations with the model for each scenario. The total population was considered to be 1.

| Parameter/initial condition | Description | Value | Scenario(s) |
|---|---|---|---|
| $\alpha$ | *Replacement of susceptible individuals* | $\sim 1.6 \times 10^{-5}$* | *0,1, 2 e 3* |
| $\phi$ | *Removal/replacement rate* | $\sim 10^{-5}$ | *0,1, 2 e 3* |
| $\beta$ | *Transmission rate* | *Variable* | *0,1, 2 e 3* |
| $\kappa$ | *The multiple of the transmissibility of $I\_a$ to $I\_s$* | *0.5* | *0,1, 2 e 3* |
| $\delta$ | *The proportion of asymptomatic infection rate* | *0.5* | *0,1, 2 e 3* |
| $\omega$ | *The incubation frequency* | *0.1923[1]* | *0,1, 2 e 3* |
| $\gamma$ | *The latency frequency* | *0.1724[1]* | *0,1, 2 e 3* |
| $\sigma$ | *Proportion of individuals in quarantine* | *Variable* | *1, 2 e 3* |
| $\nu = \sigma$ | *Proportion of individuals in isolation* | *Variable* | *2 e 3* |
| *S(0)* | *Initial proportion of susceptible individuals* | *0.8* | *0,1, 2 e 3* |
| *E(0)* | *Initial proportion of exposed individuals* | *0.0* | *0,1, 2 e 3* |
| *I_s(0)* | *Initial proportion of symptomatic individuals* | $\sim 5 \times 10^{-7}$** | *0,1, 2 e 3* |
| *I_a(0)* | *Initial proportion of asymptomatic individuals* | *0.0* | *0,1, 2 e 3* |
| *R(0)* | *Initial proportion of removed individuals* | *0.0* | *0,1, 2 e 3* |

*Portion of total initial susceptible population ($\phi S(0)$)
**Approximately 100 individuals from the entire population of Brazil.
Table 1. List of parameters and initial conditions used in the simulations.

## 3. Results and discussion.

From the results of the simulations made using our model, we were able to compare the peaks of infection in different scenarios. Fig. 1. shows the total number of the infected individuals as a function of time for the 4 different scenarios. These curves are solution of the systems of equations described above. We can identify infection peaks with a well-defined Gaussian profile. As expected, scenarios 0 and 3 do not constitute reality as they are extreme cases. Most countries should expect a scenario that may be located between 1 and 2 represented in this figure. Scenario 1, which considers the quarantine of infected individuals as a containment measure, represents a peak of infection in which just over 6% of the country's susceptible population will become infected at some point. In this scenario, the peak is reached around day 101 (just over 3 months) from the first infected case (not necessarily the first reported case) and the variance, which represents most of the time that the epidemic lasts, is approximately 38 days. A much more promising scenario is number 2. In this case, the peak represents just over 1% of the infected population and, on the other hand, this peak is attained after 224 days, from the first case. This allows the health system of any country to prepare itself to be in a position to meet the demands arising from the pandemic. The table below (Table 2) appears in a summary of this comparison.

| Scenario | Peak height | Peak position | Variance |
|---|---|---|---|
| *0* | *~11.3%* | *~71 days* | *~30 days* |
| *1* | *~6.5%* | *~102 days* | *~38 days* |
| *2* | *~1.3%* | *~224 days* | *~75 days* |
| *3* | *~0.3%* | *~431 days* | *~147 days* |

Table 2. Characteristic values of peaks of infection in different scenarios (graph in Figure 2).

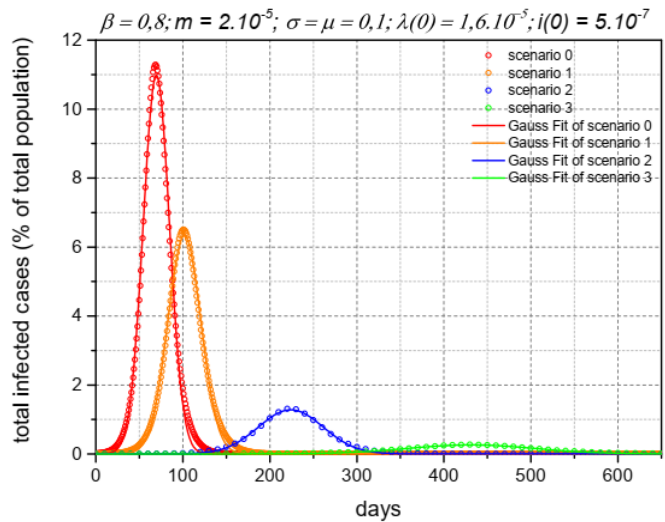

Figure 2. Total infected cases *versus* time in days, for all scenarios. The solid lines represent the Gaussian fitting. In red, the dynamic corresponding to the worst scenario is represented, while the most optimistic scenario is represented in green.

Another study carried out using our model was the comparison, within the same scenario, of the behaviours of the peaks of infection due to the variable parameters characteristic of health and social actions carried out in order to minimize the effects of the pandemic. In the graph in Figure 3, four different dynamics are represented, based on a fixed proportion of individuals who adopted the containment measures specific to this scenario. In this chart, the behaviours of peak infection are compared for unusual infection rate values. This variable can be directly related to individual health measures or at the state level, to be adopted by people during the epidemic. We studied the behaviour of four values of $\beta$, 0.95, 0.90, 0.85 and 0.80. The higher the $\beta$ value, the more likely the susceptible individual is supposed to be exposed to the virus. In this comparison we see the way the height of the peak gets smaller as $\beta$ is also smaller, representing a lesser infection. The peak position is also sensitive to this parameter, as well as the curve variance. In relation to the peak, we observed that lower values of $\beta$ guarantee a postponement in the occurrence of the peak of infection, however, the duration of the effects of the epidemic is longer.

The following table (Table 3) shows the values comparatively. Here

| $\beta$ | Peak height | Peak position | Variance |
|---|---|---|---|
| *0.95* | *~8.8%* | *~71 days* | *~31 days* |
| *0.90* | *~8.1%* | *~79 days* | *~33 days* |
| *0.85* | *~7.3%* | *~89 days* | *~35 days* |
| *0.80* | *~6.5%* | *~102 days* | *~38 days* |

Table 3. Characteristic values of peaks of infection in different scenarios (graph in Figure 3).

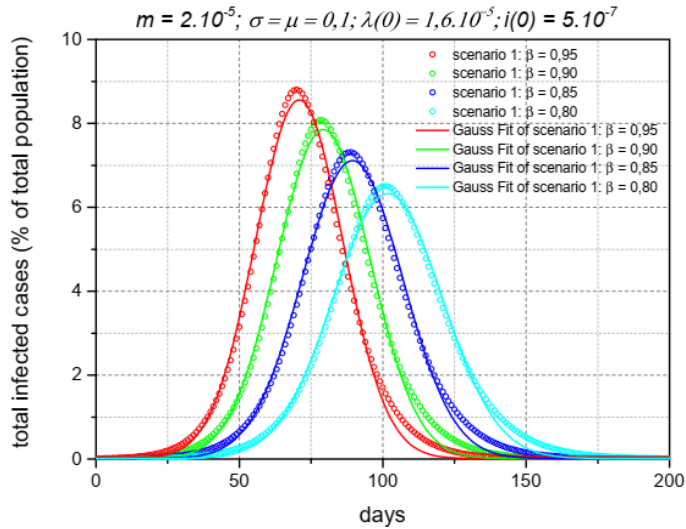

Figure 3. Total infected cases *versus* time in days, for scenario1. The solid lines represent the Gaussian fitting. Here we use $\sigma = \mu = 0.1$ and analyze several β values. The data in red represent a dynamic with a higher transmission rate. The data corresponding to the dynamics with the lowest transmission rate are shown in cyan.

The following comparison was made for the same $\beta$ value, varying the proportions of individuals in quarantine or isolation, also for a single scenario. In the graph of Figure 4, we can see the behaviour of the peaks of infection for three specific values of $\sigma$ and $\nu$. As these values are higher, this means that the containment measures for tackling the pandemic have been more effective within the specific scenario. The characteristic values of the peaks of infection can be compared in Table 4 below.

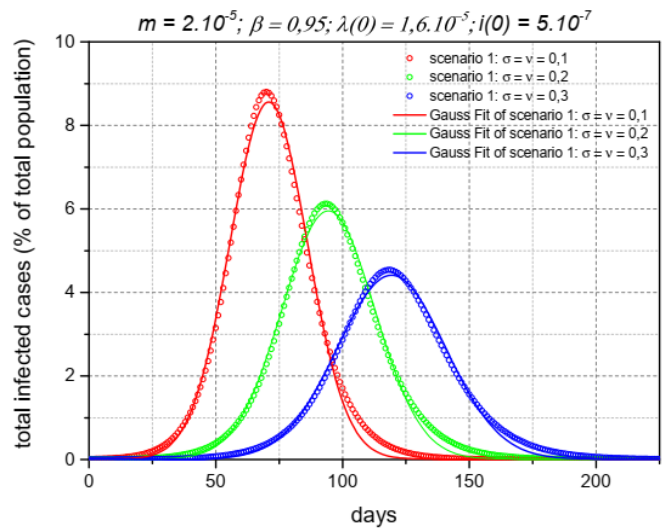

Figure 4. Total infected cases *versus* time in days, for all scenario1. The solid lines represent the Gaussian fitting. Here we use $\beta = 0.95$ and analyze various values of $\sigma = \nu$. The data in red represent a dynamic with the highest proportion of individuals adopting containment measures. The data corresponding to the dynamics with a lower proportion of people adopting containment measures are represented in blue.

| $\sigma = \nu$ | Peak height | Peak position | Variance |
|---|---|---|---|
| 0.1 | 8.8% | ~71 days | ~31 days |
| 0.2 | ~6.1% | ~94 days | ~36 days |
| 0.3 | ~4.5% | ~119 days | ~43 days |

Table 4. Characteristic values of peaks of infection in the different scenarios (graph in Figure 4).

The epidemic is growing in Brazil at a pace similar to that of other European countries. Below we display a graph similar to that of Figure 2 previously analysed. This graph in Figure 5 shows what is planned for the same four possible scenarios in relation to the cases reported in Brazil. According to the behaviour of the epidemic predicted by our model, a pessimistic scenario would lead to a maximum number of around 564000 infected individuals at the time of day 70, from the first reported case. This scenario would be impossible to sustain for our health system because the number of infected people would far exceed the capacity of available hospitals beds. A less pessimistic scenario, but still not very encouraging, would be scenario 1, with a total number of reported cases equal to approximately 257000, around the 101st day of the epidemic starting. The postponement of the peak beyond that provided in scenario 1 would be an important and decisive factor to save time and the health system to be able to prepare to face the pandemic. This scenario would be a bit worse than that estimated by ABIN days ago. A very encouraging scenario would be on the brink of scenario 2, where the maximum number of reported cases would be approximately 50000. In Table 5 we summarize the characteristic values of each peak of infection corresponding to the graph in Figure 5.

| Scenario | Reported cases | Peak position | Variance |
|---|---|---|---|
| 0 | ~564000 | ~70 days | ~30 days |
| 1 | ~257000 | ~101 days | ~37 days |
| 2 | ~50000 | ~223 days | ~74 days |
| 3 | ~13100 | ~431 days | ~147 days |

Table 5. Characteristic values of peaks of infection in different scenarios (graph in Figure 5).

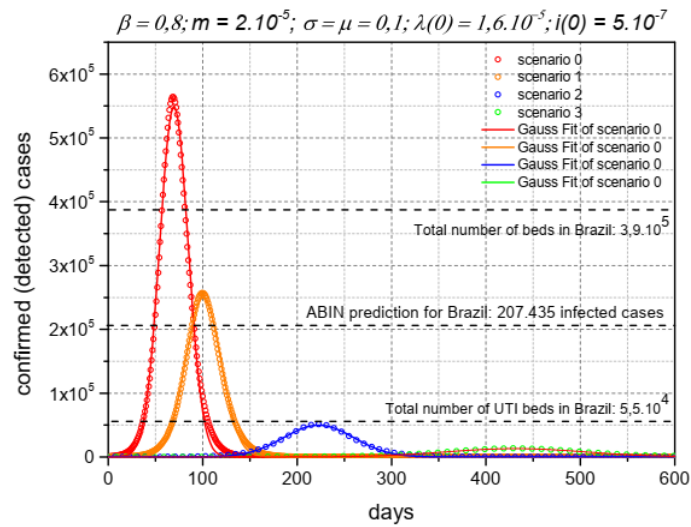

Figure 5. Confirmed infected cases versus time in days, including all scenarios. The solid lines are the Gaussian fitting. In red, the dynamic corresponding to the worst scenario is shown, while the most optimistic scenario is represented in green.

Actual data on SARS-Cov-2 infected cases have been reported by the Ministry of Health and the state departments since the beginning of the epidemic. The graph in Figure 6 shows a comparison between the dynamics of the initial infection curve for each scenario and the actual case curve in Brazil in the first 35 days. This comparison appears in a real scenario whose growth rate behaves slightly above the rate corresponding to scenario 2 of the model. Table 6 below lists the growth rates for the four scenarios and the actual reported data. This rate was obtained by graphical appreciation, based on the graphs in Figure 6, with the sole purpose of analysing these growths qualitatively.

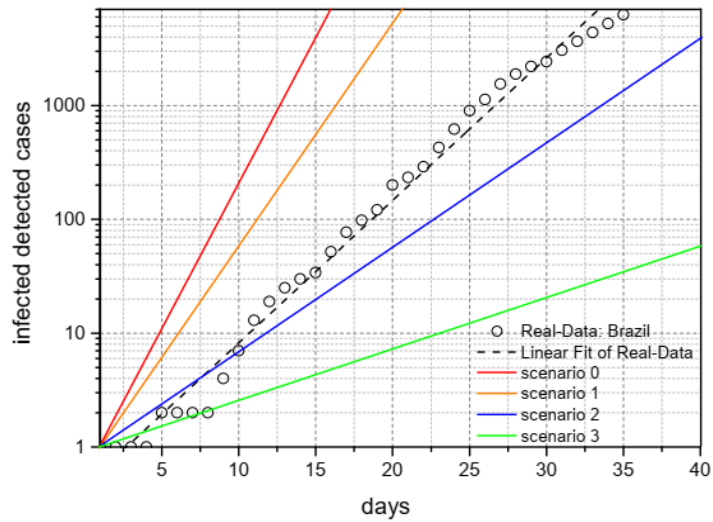

Figure 6. Representative curves of the initial exponential growth. The colours from red to green represent scenarios 0, 1, 2 and 3 respectively. The open circles are the actual data reported in Brazil during the first days after the epidemic began [8]. The dashed line is the exponential adjustment of the actual data.

| Scenario | Exponential growth rate |
|---|---|
| 0 | 0.44 |
| 1 | 0.35 |
| 2 | 0.15 |
| 3 | 0.04 |
| Real data | **0.22** |

Table 6. Growth rates of exponential curves for different scenarios compared to the actual date of cases reported in Brazil in the first 35 days of the pandemic.

Recently, interesting results have been reported showing methods that make it possible to clearly distinguish the point at which exponential growth is abandoned, depending on the total number of reported cases compared to new cases [9]. The graph in Figure 7 below shows an attempt to study precisely this behaviour, based on the results of the simulations using our model. The results shown in the graph in Figure 7 reveal the percentage that represents in each scenario the number of new cases reported in relation to the total cases. This quantity remains practically constant until the limit of the epidemic, where the proportion of new cases begins to grow more slowly, reaching until approximately 6%, before starting to decrease sharply, for the most pessimistic scenario. Such trend gives us an idea of how close we are to the end of the process. These results were obtained by simulating conditions of high infection rate and low adherence to containment measures. In Table 7 we report the proportions of new cases in relation to the total cases reported for each simulated scenario.

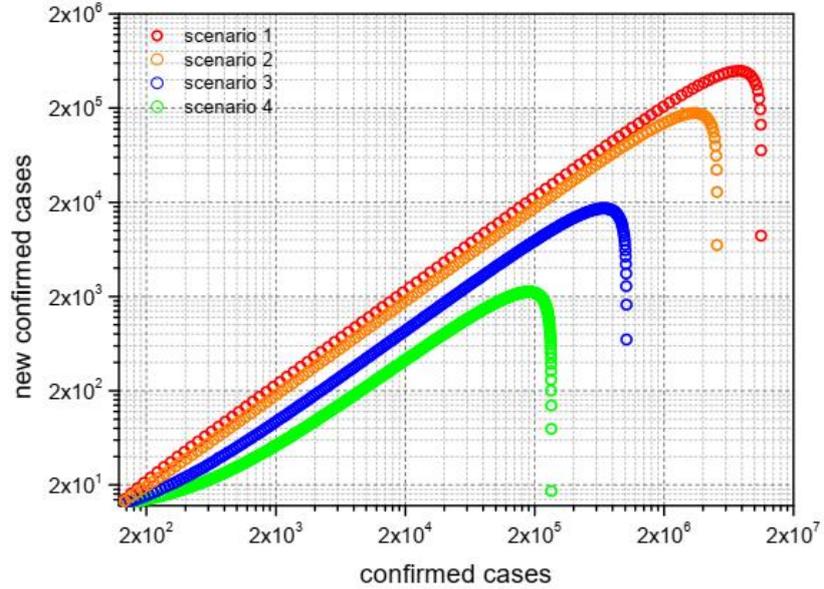

Figure 7. Curves representing the potential growth of new cases versus cases confirmed daily, derived from the model. The colours from red to green represent scenarios 0, 1, 2 and 3 respectively. The values of the parameters used in this simulation were $\beta = 0.95$ and $\sigma = \nu = 0.1$.

| Scenario | Proportion of new cases in relation to the total, at the end of the epidemic |
|---|---|
| 0 | ~6% |
| 1 | ~5% |
| 2 | ~2.5% |
| 3 | ~1.2% |

Table 7. Proportion of new cases reported in relation to the total cases reported for simulations made with $\beta = 0.95$ and $\sigma = \nu = 0.1$.

The basic reproduction number R0 is an important parameter when it comes to studying epidemics and their dynamics of infection. From the parameters of our model, including all possible scenarios, we can define R0, in each scenario, as follows:

Scenario 0:
$$R_0 = \frac{\beta(1+\kappa)+\omega}{2\gamma} \qquad (18)$$

Scenario 1:
$$R_0 = \frac{\beta(1+\kappa)+\omega}{2\gamma+\sigma} \qquad (19)$$

Scenario 2:
$$R_0 = \frac{\beta(1+\kappa)+\omega}{2\gamma+\sigma+\nu} \qquad (20)$$

Scenario 3:

$$R_0 = \frac{\beta(1+\kappa)+\omega}{2(\gamma+\nu)+\sigma} \tag{21}$$

Estimates of $R_0$ were made in all scenarios and for all values of $\sigma$ and $\nu$. The values of the basic reproduction numbers in each simulated scenario are shown in Table 8. The transmission rate used was 0.8.

| Scenario | $\sigma$ | $\nu$ | $R_0$ |
| --- | --- | --- | --- |
| 0 | 0 | 0 | 4.04 |
| 1 | 0.1 | 0 | 3.13 |
| 1 | 0.2 | 0 | 2.25 |
| 1 | 0.3 | 0 | **2.16** |
| 2 | 0.1 | 0.1 | 2.55 |
| 2 | 0.2 | 0.2 | 1.87 |
| 2 | 0.3 | 0.3 | 1.47 |
| 3 | 0.1 | 0.1 | **2.16** |
| 3 | 0.2 | 0.2 | 1.47 |
| 3 | 0.3 | 0.3 | 1.12 |

Table 8. Basic reproduction number values in each scenario and for each value of $\sigma$ and $\nu$.

From these basic reproduction number results obtained in the simulations, we can see that in terms of R0 there are scenarios that, depending on the values of the characteristic parameters, can be equivalent. For the values used in the simulations, scenario 1, using σ = 0.3, is equivalent to scenario 3, with σ = ν = 0.1. This means that more promising scenarios in terms of the number of measures may not be really promising if adherence to these measures is not high or if these measures are not efficient.

## 4. Conclusions.

The studies carried out based on the results obtained from the simulations show, in the first place, the importance of containment measures to slow the spread of the epidemic. These measures guarantee, the flattening of the peak of infection and the delay in the moment of peak occurrence, two positive consequences of the adoption of these measures. The advance of the epidemic in Brazil shows intermediate characteristics that fit between scenarios 1 and 2, which is not a very promising situation. In these scenarios, the containment measures are reduced to the quarantine of infected individuals and the isolation of exposed people. The containment measures do not seem to be being efficient and this may be the result of poor application, poor adherence or the delay in being more strictly enacted by the competent authorities. It is also demonstrated that the reproducibility of the virus depends on each specific scenario. Finally, we see that it is very important, in order to more accurately determine the final moment of the epidemic, to inspect as accurately as possible the number of new cases daily in relation to the total reported cases.

I thank Fernando de León Pérez for the critical Reading of the manuscript.

## 5. References.